# Theoretical modeling of critical temperature increase in metamaterial superconductors

Igor I. Smolyaninov [1] and Vera N. Smolyaninova [2]

*[1]Department of Electrical and Computer Engineering, University of Maryland, College Park, MD 20742, USA*

*[2]Department of Physics Astronomy and Geosciences, Towson University,*

*8000 York Rd., Towson, MD 21252, USA*

**Recent experiments have demonstrated that the metamaterial approach is capable of drastic increase of the critical temperature $T_c$ of epsilon near zero (ENZ) metamaterial superconductors. For example, tripling of the critical temperature has been observed in Al-Al$_2$O$_3$ ENZ core-shell metamaterials. Here, we perform theoretical modelling of $T_c$ increase in metamaterial superconductors based on the Maxwell-Garnett approximation of their dielectric response function. Good agreement is demonstrated between theoretical modelling and experimental results in both aluminium and tin-based metamaterials. Taking advantage of the demonstrated success of this model, the critical temperature of hypothetic niobium, MgB$_2$ and H$_2$S-based metamaterial superconductors is evaluated. The MgB$_2$-based metamaterial superconductors are projected to reach the liquid nitrogen temperature range. In the case of an H$_2$S-based metamaterial $T_c$ appears to reach ~250 K.**

Our recent theoretical [1,2] and experimental [3,4] work demonstrated that many tools developed in electromagnetic metamaterial research may be successfully used to engineer artificial metamaterial superconductors having considerably improved superconducting properties. This deep and non-trivial connection between the fields of electromagnetic metamaterials and superconductivity research stems from the fact that superconducting properties of a material, such as electron-electron pairing interaction, the superconducting critical temperature $T_c$, etc. may be expressed via the effective dielectric response function $\varepsilon_{eff}(q,\omega)$ of the material [5]. For example, considerable enhancement of attractive electron-electron interaction may be expected in such actively studied metamaterial scenarios as epsilon near zero (ENZ) [6] and hyperbolic metamaterials [7] since in both cases $\varepsilon_{eff}(q,\omega)$ may become small and negative in substantial portions or the four-momentum $(q,\omega)$ space. Such an effective dielectric response-based macroscopic electrodynamics description is valid if the material may be considered as a homogeneous medium on the spatial scales below the superconducting coherence length. The metamaterial superconductor approach takes advantage of the recent progress in plasmonics and electromagnetic metamaterials to engineer an artificial medium or "metamaterial", so that its effective dielectric response function $\varepsilon_{eff}(q,\omega)$ conforms to almost any desired behaviour. It appears natural to use this newly found freedom to engineer and maximize the electron pairing interaction in such an artificial superconductor via engineering its dielectric response function $\varepsilon_{eff}(q,\omega)$.

The most striking example of successful metamaterial superconductor engineering was recent observation of tripling of the critical temperature $T_c$ in Al-Al$_2$O$_3$ epsilon near zero (ENZ) core-shell metamaterial superconductors compared to bulk aluminium [4]. However, both this result and the previous demonstration of $T_c$ increase in random ENZ mixtures [3] were based on qualitative theoretical guidelines, while numerical tools enabling quantitative prediction of $T_c$ increase in a metamaterial superconductor were missing. Here we perform quantitative theoretical modelling of $T_c$ increase in ENZ metamaterial superconductors based on the Maxwell-Garnett approximation of their dielectric response function. Good agreement is demonstrated between theoretical modelling and experimental results. Taking advantage of the demonstrated success of this model, the critical temperature of hypothetic niobium, $MgB_2$ and $H_2S$-based metamaterial superconductors is evaluated. In the case of $H_2S$-based metamaterial $T_c$ appears to reach ~250 K range.

The starting point of our model is the paper by Kirzhnits *et al.* [5], which demonstrated that within the framework of macroscopic electrodynamics the electron-electron interaction in a superconductor may be expressed in the form of an effective Coulomb potential

$$V(\vec{q}, \omega) = \frac{4\pi e^2}{q^2 \varepsilon_{eff}(\vec{q}, \omega)},$$  (1)

where $V_C = 4\pi e^2/q^2$ is the Fourier-transformed Coulomb potential in vacuum, and $\varepsilon_{eff}(q, \omega)$ is the linear dielectric response function of the superconductor treated as an effective medium. Following [8], a simplified dielectric response function of a metal may be written as

$$\varepsilon_m(q, \omega) = \left(1 - \frac{\omega_p^2}{\omega^2 + i\omega\Gamma - \omega_p^2 q^2/k^2}\right)\left(1 - \frac{\Omega_1^2(q)}{\omega^2 + i\omega\Gamma_1}\right)\ldots\left(1 - \frac{\Omega_n^2(q)}{\omega^2 + i\omega\Gamma_n}\right)$$  (2)

where $\omega_p$ is the plasma frequency, $k$ is the inverse Thomas-Fermi radius, $\Omega_n(q)$ are dispersion laws of various phonon modes, and $\Gamma_n$ are the corresponding damping rates. Zeroes of the dielectric response function of the bulk metal (which correspond to various bosonic modes) maximize electron-electron pairing interaction given by Eq. (1). In the discussion below we will limit our consideration to the behaviour of $\varepsilon_m(q, \omega)$ near $\omega = \Omega_1(q)$ in the vicinity of the Fermi surface, so that a more complicated representation of $\varepsilon_m(q,\omega)$ by the Lindhard function, which accurately represents $\varepsilon_m(q,\omega)$ in the $q \to 0$ and $\omega \to 0$ limit [8] does not need to be used. We are going to use only the fact that $\varepsilon_m$ changes sign near $\omega = \Omega_1(q)$ by passing through zero. Expression for $\varepsilon_m(q,\omega)$ given by Eq.(2) does exhibit this behavior.

As summarized in [9], the critical temperature of a superconductor in the weak coupling limit is typically calculated as

$$T_c = \theta \; \exp\left(-\frac{1}{\lambda_{eff}}\right),$$
(3)

where $\theta$ is the characteristic temperature for a bosonic mode mediating electron pairing (such as the Debye temperature $\theta_D$ in the standard BCS theory, or $\theta_{ex} = \hbar\omega_{ex}/k$ or $\theta_{pl} = \hbar\omega_{pl}/k$ in the theoretical proposals which suggest excitonic or plasmonic mediation of the electron pairing interaction), and $\lambda_{eff}$ is the dimensionless coupling constant defined by $V(q,\omega) = V_C(q) \; \varepsilon^{-1}(q,\omega)$ and the density of states $\nu$ (see for example [10]):

$$\lambda_{eff} = -\frac{2}{\pi}\nu\int\limits_0^\infty \frac{d\omega}{\omega}\left\langle V_C(q)\,\mathrm{Im}\,\varepsilon^{-1}\!\left(\vec{q},\omega\right)\right\rangle,$$
(4)

where $V_C$ is the unscreened Coulomb repulsion, and the angle brackets denote average over the Fermi surface.

Compared to the bulk metal, zeroes of the effective dielectric response function $\varepsilon_{eff}(q,\omega)$ of the metal-dielectric metamaterial are observed at shifted positions compared to the zeroes of $\varepsilon_m(q,\omega)$ [2], and additional zeros may also appear. According to the Maxwell-Garnett approximation [11], mixing of nanoparticles of a superconducting "matrix" with dielectric "inclusions" (described by the dielectric constants $\varepsilon_m$ and $\varepsilon_d$, respectively) results in the effective medium with a dielectric constant $\varepsilon_{eff}$, which may be obtained as

$$\left(\frac{\varepsilon_{eff} - \varepsilon_m}{\varepsilon_{eff} + 2\varepsilon_m}\right) = (1-n)\left(\frac{\varepsilon_d - \varepsilon_m}{\varepsilon_d + 2\varepsilon_m}\right), \tag{5}$$

where $n$ is the volume fraction of metal ($0 \le n \le 1$). The explicit expression for $\varepsilon_{eff}$ may be written as

$$\varepsilon_{eff} = \frac{\varepsilon_m\big((3-2n)\varepsilon_d + 2n\varepsilon_m\big)}{\big(n\varepsilon_d + (3-n)\varepsilon_m\big)}, \tag{6}$$

and it is easy to verify that

$$\varepsilon_{eff}^{-1} = \frac{n}{(3-2n)}\frac{1}{\varepsilon_m} + \frac{9(1-n)}{2n(3-2n)}\frac{1}{(\varepsilon_m + (3-2n)\varepsilon_d/2n)} \tag{7}$$

For a given value of the metal volume fraction $n$, the ENZ conditions ($\varepsilon_{eff} \approx 0$) may be obtained around $\varepsilon_m \approx 0$ and around

$$\varepsilon_m \approx -\frac{3-2n}{2n}\varepsilon_d \tag{8}$$

Since the absolute value of $\varepsilon_m$ is limited by some value $\varepsilon_{m,max}$ (see Eq. (2)), the latter zero of $\varepsilon_{eff}$ disappears at small $n$ as $n \to 0$ at some critical value of the volume fraction $n = n_{cr} = 1.5\varepsilon_d/\varepsilon_{m,max}$. Let us evaluate $\mathrm{Im}(\varepsilon_{eff}^{-1}(q,\omega))$ near this zero of $\varepsilon_{eff}$ at $n > n_{cr}$. Based on Eqs. (7),

$$\mathrm{Im}\big(\varepsilon_{eff}^{-1}\big) \approx \frac{9(1-n)}{2n(3-2n)\varepsilon''_m} \tag{9}$$

where $\varepsilon_m' = \mathrm{Re}\,\varepsilon_m$ and $\varepsilon_m'' = \mathrm{Im}\,\varepsilon_m$. Assuming that $\nu \sim n$ (which is justified by the fact that there are no free charges in the dielectric phase of the metamaterial) and using Eq. (4), we may obtain the resulting expression for $\lambda_{eff}$ as a function of $\lambda_m$ and $n$:

$$\lambda_{eff} \approx \frac{9(1-n)}{2(3-2n)}\lambda_m, \tag{10}$$

where $\lambda_m$ is also determined by Eq.(4) in the limit $\varepsilon \to \varepsilon_m$, and we have assumed the same magnitudes of $\varepsilon_m''$ for the metamaterial zero described by Eq. (8) and for the $\varepsilon_m \approx 0$ conditions. This assumption will be re-evaluated later in the paper (see Eq. (16)) and confronted with experimental data for aluminium-based ENZ metamaterial, which indicate that $\varepsilon_m''$ changes by ~11%. Fig. 1a shows the predicted behaviour of $\lambda_{eff}/\lambda_m$ as a function of $n$. Based on this prediction, we may expect enhancement of superconducting properties of the metal-dielectric metamaterial in the $n_{cr} < n < 0.6$ range. For comparison, Fig. 1a also shows the behaviour of $\lambda_{eff}$ near the $\varepsilon_m \approx 0$ pole of the inverse dielectric response function of the metamaterial. According to Eq. (6), near $\varepsilon_m \approx 0$ the effective dielectric response function of the metamaterial behaves as

$$\varepsilon_{eff} \approx \frac{3 - 2n}{n} \varepsilon_m \qquad (11)$$

Therefore, near this pole

$$\lambda_{eff} \approx \frac{n^2}{(3 - 2n)} \lambda_m \qquad (12)$$

Let us apply the obtained simple estimates to the case of Al-Al$_2$O$_3$ core-shell metamaterial studied in [4]. Assuming the known values $Tc_{bulk} = 1.2$ K and $\theta_D = 428$ K of bulk aluminium [8], Eq. (3) results in $\lambda_m = 0.17$, which corresponds to the weak coupling limit. In order to make the mechanism behind the enhancement of $T_c$ in the Al-Al$_2$O$_3$ core-shell metamaterial superconductor abundantly clear, let us plot the hypothetic values of $T_c$ as a function of metal volume fraction $n$, which would originate from either Eq. (10) or Eq. (12) in the absence of each other. The corresponding values calculated as

$$T_c = T_{Cbulk} \exp\left( \frac{1}{\lambda_m} - \frac{1}{\lambda_{eff}} \right) \qquad (13)$$

are shown in Fig. 1b. The vertical dashed line corresponds to the assumed value of $n_{cr}$. The experimentally measured data points from Ref. 4 are shown for comparison on the same plot. The match between the experimentally measured values of enhanced $T_c$ and the theoretical curve obtained based on Eq. (10) is quite impressive, given the fact that Eqs. (10) and (13) do not contain any free parameters. Such a good match unambiguously identifies the metamaterial enhancement as the physical mechanism of critical temperature tripling in the Al-Al$_2$O$_3$ core-shell metamaterial superconductor.

Before continuing our theoretical analysis, let us briefly describe how the experimental $T_c(n)$ behaviour may be recovered from the data presented in Ref. 4 (since such an analysis has not been presented previously). The 18 nm diameter Al nanoparticles for these experiments were acquired from the US Research Nanomaterials, Inc. Upon exposure to the ambient conditions a ~ 2 nm thick Al$_2$O$_3$ shell is known to form on the aluminium nanoparticle surface [12], which is comparable to the 9 nm radius of the original Al nanoparticle. Further aluminum oxidation may also be achieved by heating the nanoparticles in air. The resulting core-shell Al$_2$O$_3$-Al nanoparticles were compressed into macroscopic, ~ 1 cm diameter, ~ 0.5 mm thick test pellets using a hydraulic press. While the $T_c$ of various Al-Al$_2$O$_3$ core-shell metamaterials was determined directly via the onset of diamagnetism for samples with different degrees of oxidation using a MPMS SQUID magnetometer (see Fig. 2a), the volume fraction of aluminium $n$ in the metamaterial was determined based on the measured FTIR reflectivity spectra of the metamaterial (Fig. 2b). Surface roughness of the metamaterial samples complicates quantitative analysis of such spectra due to wavelength-dependent scattering. This difficulty has been overcome by careful analysis of the measured FTIR reflectivity spectra in the vicinity of phonon-polariton resonance in Al$_2$O$_3$ [13]. Due to this resonance, pure Al$_2$O$_3$ behaves as an almost perfect absorber at 10.3 μm wavelength, while it behaves as an almost perfect metal (and therefore reflects almost 100% light) at 12 μm. Since these wavelengths are very close to each other, the effect of wavelength-dependent scattering should be almost the same, while reflectivity of aluminium at 12 μm is also almost perfect. Therefore, this phonon-polariton resonance feature may be used to accurately determine volume fraction of Al$_2$O$_3$ in the metamaterial. In Fig .2b all the measured FTIR curves were normalized to 100% reflectivity at 12 μm wavelength to negate the effect of scattering. As a result, the normalized reflectivity of the metamaterial samples at 10.3 μm (where Al$_2$O$_3$ behaves as an almost perfect absorber) indicates the volume fraction of aluminium in the metamaterial sample. The so obtained $T_c(n)$ data points were plotted in Fig. 1b and compared with the theoretical curves.

While theoretical plots in Fig. 1b unambiguously identify the physical mechanism of $T_c$ enhancement in the Al-Al$_2$O$_3$ core-shell metamaterial superconductor, it must be understood that a complete theory should take into account simultaneous contributions of both poles of Eq. (7) to $\lambda_{\text{eff}}$ of the metamaterial. It is also clear that according to Eq. (2) the metamaterial pole occurs at a slightly different frequency compared to the $\varepsilon_m \approx 0$ pole, and therefore it must have slightly different value of $\varepsilon_m'' = \text{Im}\,\varepsilon_m$ compared to the value of $\varepsilon_m''$ for pure aluminium at $\text{Re}\,\varepsilon_m \approx 0$. In order to obtain more precise values, let us consider the behaviour of $\varepsilon_m$ in more detail. Since

$$\left(1 - \frac{\Omega_1^2(q)}{\omega^2 + i\omega\Gamma_1}\right) = \frac{(\omega + \Omega_1)(\omega - \Omega_1) + i\omega\Gamma_1}{\omega^2 + i\omega\Gamma_1}, \qquad (14)$$

let us assume that near $\omega = \Omega_1(q)$ Eq. (2) may be approximated as

$$\varepsilon_m \approx -E_m(\omega - \Omega_1) + i\varepsilon_m''(\omega) \quad, \qquad (15)$$

where $E_m$ is a positive constant. Based on Eq. (8), the corresponding frequency of the "metamaterial" pole is

$$\omega = \Omega_1 + \frac{(3-2n)\varepsilon_d}{2nE_m},$$  (16)

which is slightly higher than $\Omega_1$, and therefore a slightly larger value of $\mathrm{Im}(\varepsilon_m) = \varepsilon_m''\left(\Omega_1 + (3-2n)\varepsilon_d / 2nE_m\right)$ may be expected at the metamaterial pole at this higher frequency. As a result, by taking into account simultaneous contributions from both poles (by adding the contributions to $\lambda_{eff}$ from the "metal" and the "metamaterial" poles given by Eqs.(10) and (12), respectively), the final expression for $\lambda_{eff}$ is

$$\lambda_{eff} \approx \frac{n^2}{(3-2n)}\lambda_m + \frac{9(1-n)}{2(3-2n)}\frac{\varepsilon_m''}{\varepsilon_{mm}''}\lambda_m = \lambda_m\left(\frac{n^2}{(3-2n)} + \frac{9(1-n)\alpha}{2(3-2n)}\right),$$  (17)

where $\alpha = \varepsilon_m'' / \varepsilon_{mm}'' < 1$ is determined by the dispersion of $\varepsilon_m''$. Substitution of Eq. (17) into Eq. (13) produces the following final expression for the critical temperature of the metamaterial:

$$T_c = T_{Cbulk}\exp\left(\frac{1}{\lambda_m}\left(1 - \frac{1}{\left(\frac{n^2}{(3-2n)} + \frac{9(1-n)\alpha}{2(3-2n)}\right)}\right)\right) \text{ at } n > n_{cr},$$  (18)

which now has a single fitting parameter $\alpha$. We will consider it as a free parameter of the model, since we are not aware of any experimental measurements of dispersion of $\mathrm{Im}\,\varepsilon_m$ for aluminum. Note that Eq. (18) depends on $\varepsilon_d$ only via $n_{cr}$ and $\alpha$. The calculated behavior of $T_c$ as a function of $n$ is presented in Fig.3a. The agreement appears to be very good. The best fit to experimental metamaterial data is obtained at $\alpha = 0.89$. We anticipate that the same theory may explain the increase of $T_c$ in granular aluminum films [14,15]. However, we are not aware of accurate measurements of Al/Al$_2$O$_3$ volume fraction in these samples. We must emphasize that our model is quite insensitive to the particular choice of functional form of $\varepsilon_m(q,\omega)$ in the broad $(q,\omega)$ range. In our derivations we are only using the fact that $\varepsilon_m$ changes sign near $\omega = \Omega_1(q)$ by passing through zero, which is described by Eq.(15). A simplified dielectric response function of a metal written in the form of Eq.(2) [8] does exhibit such behaviour. However, other expressions for $\varepsilon_m(q,\omega)$, such as the Lindhard function [8], or the dielectric function of a BCS superconductor derived by Mattis and Bardeen [16] may be used in our derivation with equal success.

The apparent success of such a simple theoretical description in the case of Al-Al$_2$O$_3$ core-shell metamaterial superconductors prompted us to apply the same theory to the tin-based ENZ metamaterials studied in [3]. Assuming the known values $Tc_{bulk}$=3.7K and $\theta_D$ = 200 K of bulk tin [8], Eq. (3) results in $\lambda_m$ = 0.25, which also corresponds to the weak coupling limit. The calculated behavior of $T_c$ as a function of $n$ is presented in Fig. 3b, which also shows experimental data points measured in [3]. The agreement also appears to be very good, given the spatial inhomogeneity of the fabricated metamaterials [3] (note also the difference in temperature scales in Figs. 3a and 3b). The best fit to experimental metamaterial data is obtained at $\alpha$ = 0.83. It appears that a smaller increase in metamaterial $T_c$ in this case is due to larger value of permittivity $\varepsilon_d$ of the dielectric component of the metamaterial, which according to Eq. (8) limits the range $n_{cr} < n < 0.6$ where the metamaterial $T_c$ enhancement may occur. Larger value of $\varepsilon_d$ also leads to smaller value of $\alpha$ (see Eq. (16)), which also reduces the metamaterial pole contribution to $\lambda_{eff}$.

We should also note that the appearance of an additional "metamaterial" pole in the Maxwell-Garnett expression (Eq. (5)) for a metal-dielectric mixture does not rely on any particular spatial scale, as long as it is smaller than the coherence length. In fact, the Maxwell-Garnett approximation may be considered as a particular case of the Clausius–Mossotti relation [11], which traces the dielectric constant of a mixture to the dielectric constants of its constituents, and it is only sensitive to the volume fractions of the constituents. Therefore, the same concept is supposed to be applicable at much smaller spatial scales compared to the two metamaterial cases studied in Refs. 3 and 4. We anticipate that similar $T_c$ enhancements may be observed in metamaterials based on other higher temperature superconductors, which have smaller coherence length, such as niobium and $MgB_2$.

First, let us consider the case of niobium, which has $Tc_{bulk} = 9.2$ K and $\theta_D = 275$K in the bulk form [8]. Enhancing the superconducting properties of niobium is a very important task, since niobium alloys, such as $Nb_3Sn$ and NbTi are widely used in superconducting cables and magnets. Based on Eq. (3) niobium has $\lambda_m = 0.29$, which also corresponds to the weak coupling limit. The coherence length of niobium is $\xi = 38$ nm [8], which would complicate nanofabrication requirements for a niobium-based metamaterial. However, given the current state of nanolithography, which operates on a 14 nm node [17], these requirements still look quite realistic. Alternatively, smaller than 38 nm diameter nanoparticles would need to be used in nanoparticle-based ENZ metamaterial geometries.

The calculated behavior of $T_c$ of a niobium-based ENZ metamaterial as a function of $n$ is presented in Fig. 4a for different values of the $\alpha$ ratio in Eq. (18). The vertical dashed line in the plot corresponds to the same value of $n_{cr}$ as in the $Al$-$Al_2O_3$ core-shell metamaterial superconductor. It is interesting to note that the general range of predicted $T_c$ enhancement matches well with the observed enhancement of $T_c$ in various niobium alloys [8]. The corresponding experimental data point for $Nb_3Sn$ is shown in the same plot for comparison (it is assumed that $n = 0.75$ for Nb$_3$Sn). As we have mentioned above, our model is based on the Maxwell-Garnett approximation, which traces the dielectric constant of a mixture to the dielectric constants of its constituents. The spatial scale of mixing is not important within the scope of this model. Therefore, our model may be applicable to some alloys. We should note that the additional "metamaterial" pole described by Eq. (8) may also be present in the case of a mixture of a normal metal with the superconductor. However, unlike the previously considered case of a dielectric mixed with a superconductor, the additional "metamaterial" pole is observed at $\omega < \Omega_1$ where the dielectric constant of the superconductor $\varepsilon_s$ is positive (see Eq. (15)), while the dielectric constant of the normal metal $\varepsilon_m$ is negative (it is assumed that the critical temperature of the normal metal is lower than the critical temperature of the superconductor). Indeed, under these conditions an additional pole in the inverse effective dielectric response function of the metamaterial does appear in (Eq. (7)), and similar to Eq. (8), it is defined as

$$\varepsilon_s \approx -\frac{3-2n}{2n}\varepsilon_m \qquad (19)$$

This pole will result in the enhanced $T_c$ of the metamaterial (or alloy), which may be calculated using the same Eq. (18). However, since this additional pole occurs at $\omega < \Omega_1$, it is expected that Im$\varepsilon_s$ will be smaller at this frequency, resulting in factor $\alpha$ being slightly larger than 1. This conclusion agrees well with the position of Nb$_3$Sn data point on Fig.4a.

Next, let us consider the case of MgB$_2$, which has Tc$_{bulk}$ = 39 K and $\theta_D$ = 920 K in the bulk form [18]. The coherence length of MgB$_2$ remains relatively large: the $\pi$ and $\sigma$ bands of electrons have been found to have two different coherence lengths, 51 nm and 13 nm [19]. Both values are large enough to allow metamaterial fabrication, at least in principle. Based on Eq. (3) $MgB_2$ has $\lambda_m = 0.32$, which still remains within the scope of the weak coupling limit. The calculated behavior of $T_c$ of a $MgB_2$-based ENZ metamaterial as a function of $n$ is presented in Fig.4b for different values of the $\alpha$ ratio in Eq. (18). Similar to Fig. 4a, the vertical dashed line in the plot corresponds to the same value of $n_{cr}$ as in the $Al$-$Al_2O_3$ core-shell metamaterial superconductor. It appears that the critical temperature of $MgB_2$-based ENZ metamaterial would probably fall in the liquid nitrogen temperature range. A good choice of the dielectric component of such a metamaterial could be diamond: 5 nm diameter diamond nanoparticles are available commercially – see for example Ref. 3.

Unfortunately, superconductors having progressively higher $T_c$ are not well described by Eq. (3), which is valid in the weak coupling limit only. So it is interesting to evaluate what kind of metamaterial enhancement may be expected in such system as $H_2S$, which superconducts at 203 K at very high pressure [20]. While simple theoretical description developed in our work may not be directly applicable to this case as far as $T_c$ calculations are concerned (because of the very large values of $\theta_D \sim 1200$ K and $\lambda \sim 2$ in this material [21]) adding a suitable dielectric or normal metal to $H_2S$ will still result in an additional "metamaterial" pole in its inverse dielectric response function. Eqs.(5- 9) will remain perfectly applicable in this case. Therefore, adding such a pole will enhance superconducting properties of $H_2S$. Based on Eq. (7) we may predict that the enhancement of superconducting properties will be observed at $n \sim 0.6$, as it also happened in the cases of aluminum and tin-based metamaterials considered in this paper. As far as the choice of a dielectric is concerned, based on Eqs. (5-9) it appears that readily available nanoparticles of diamond (off the shelf 5 nm diameter diamond nanoparticles were used in [3]) could be quite suitable, since diamond has rather low dielectric constant $\varepsilon_{dia} \sim 5.6$, which stays almost constant from the visible to RF frequency range. Low permittivity of the dielectric component ensures smaller value of $n_{cr}$ defined by Eq.(8), and hence larger metamaterial enhancement. It is not inconceivable that 5 nm diameter diamond nanoparticles could be added to $H_2S$ in the experimental chamber used in [20].

Unfortunately, in the strong coupling limit the expected $T_c$ enhancement may not be as drastic as in the weak coupling limit described by Eq. (3). In the $\lambda \gg 1$ limit of the Eliashberg theory the critical temperature of the superconductor may be obtained as [22]

$$T_c = 0.183\theta_D\sqrt{\lambda} \qquad (19)$$

Therefore, enhancement of $\lambda$ by the largest possible factor of 1.5 calculated based on the Maxwell-Garnett theory for the effective dielectric response function of the metamaterial (see Fig.1a) will lead to $T_c$ enhancement by factor of $(1.5)^{1/2}$, which means that a critical temperature range $\sim 250$ K (or $\sim$ -20$^o$ C) may potentially be achieved in a $H_2S$ based metamaterial superconductor. Such a development would still be of sufficient interest, since this would constitute an almost room temperature superconductivity.

In conclusion, we have performed theoretical modelling of $T_c$ increase in metamaterial superconductors based on the Maxwell-Garnett approximation of their dielectric response function. Good agreement is demonstrated between theoretical modelling and experimental results in both aluminum [4] and tin-based [3] metamaterials. Taking advantage of the demonstrated success of this model, the critical temperature of hypothetic niobium, $MgB_2$ and $H_2S$-based metamaterial superconductors has been evaluated. The $MgB_2$-based metamaterial superconductors are projected to reach the liquid nitrogen temperature range. In the case of an $H_2S$-based metamaterial $T_c$ appears to reach ~250 K. An alternative metamaterial scenario, which involves either artificial or natural hyperbolic metamaterials, such as BSCCO [23] will be considered elsewhere. However, it is already clear that we may now answer the question, which we have asked two years ago [1]: is there a metamaterial route to high temperature superconductivity? Based on the theoretical analysis reported here, the answer appears to be positive.

This work was supported in part by NSF grant DMR-1104676.

**Figure Captions**

**Figure 1**. (a) Predicted behaviour of $\lambda_{eff}/\lambda_m$ as a function of metal volume fraction $n$ in a metal-dielectric metamaterial. (b) Plots of the hypothetic values of $T_c$ as a function of metal volume fraction $n$, which would originate from either Eq. (10) or Eq. (12) in the absence of each other. The experimentally measured data points from [4] are shown for comparison on the same plot. The vertical dashed line corresponds to the assumed value of $n_{cr}$. Note that the theoretical curves contain no free parameters.

**Figure 2**. (a) Temperature dependence of zero field cooled magnetization per unit mass for several Al-Al$_2$O$_3$ core-shell metamaterial samples with various degrees of oxidation measured in magnetic field of 10 G. The highest onset of superconductivity at ~ 3.9 K is 3.25 times larger than $T_c$ = 1.2 K of bulk aluminium [4]. (b) Corresponding FTIR reflectivity spectra of the core-shell metamaterial samples in the region of phonon-polariton resonance in Al$_2$O$_3$. All curves are normalized to 100% reflectivity at 12 μm wavelength. The normalized reflectivity of the metamaterial samples at 10.3 μm (where Al$_2$O$_3$ behaves as an almost perfect absorber) indicates the volume fraction of aluminium in the metamaterial sample.

**Figure 3**. (a) Theoretical plot of $T_c$ versus volume fraction $n$ of aluminium in the Al-Al$_2$O$_3$ core-shell metamaterial calculated using Eq. (18). The best fit to experimental data is obtained at $\alpha$ = 0.89. (b) Theoretical plot of $T_c$ versus volume fraction $n$ of tin in the tin-BaTiO$_3$ ENZ metamaterial [3] calculated using Eq. (18). The best fit to experimental data is obtained at $\alpha$ = 0.83. The experimental data points are taken from Ref. 3.

**Figure 4**. (a) Theoretical plots of $T_c$ versus volume fraction $n$ of niobium in a hypothetic niobium-based ENZ metamaterial calculated using Eq. (18) for different values of the $\alpha$ ratio. The experimental data points for bulk Nb and Nb$_3$Sn are shown in the same plot

for comparison. (b) Theoretical plots of $T_c$ versus volume fraction $n$ of $MgB_2$ in a hypothetic $MgB_2$-based ENZ metamaterial calculated using Eq. (18) for different values of the $\alpha$ ratio. In both cases the vertical dashed lines correspond to the same value of $n_{cr}$ as in the Al-$Al_2O_3$ core-shell metamaterial superconductor.

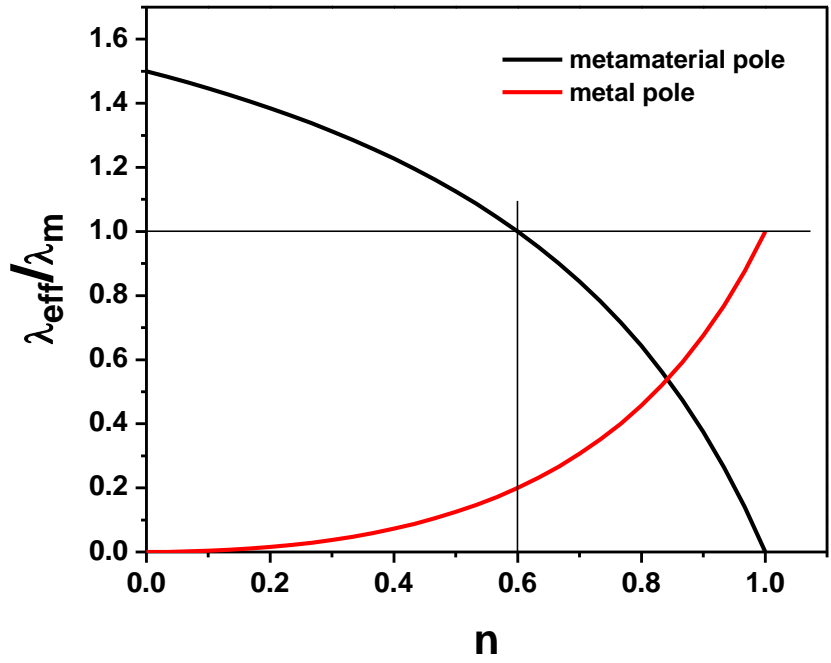

(a)

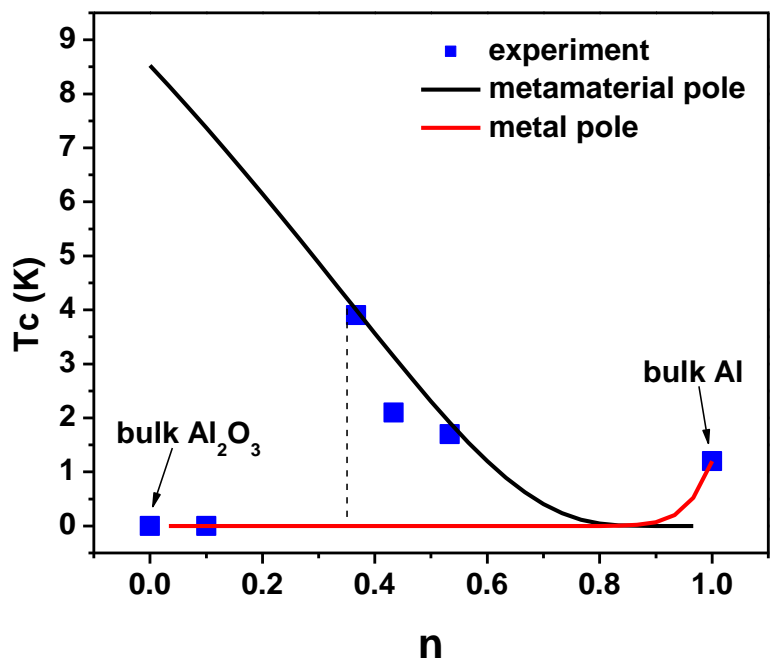

(b)

Fig. 1

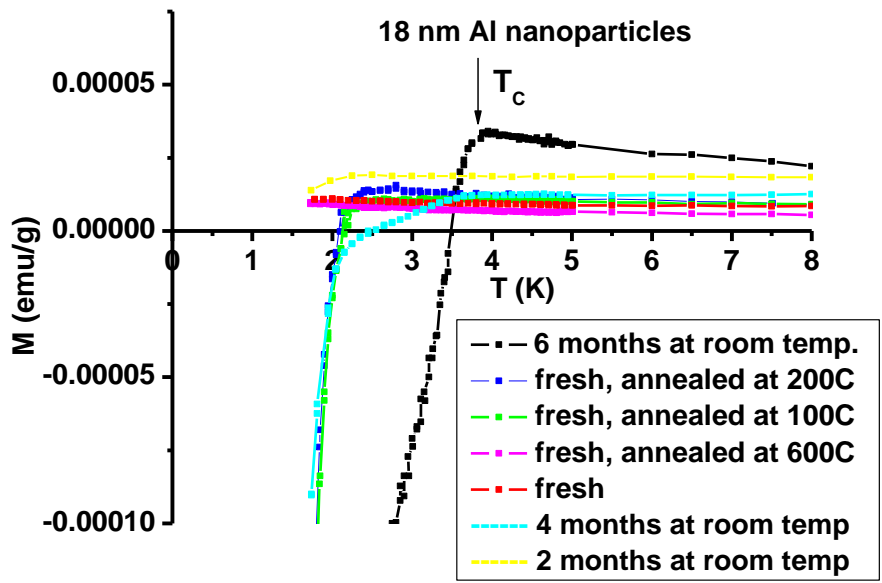

(a)

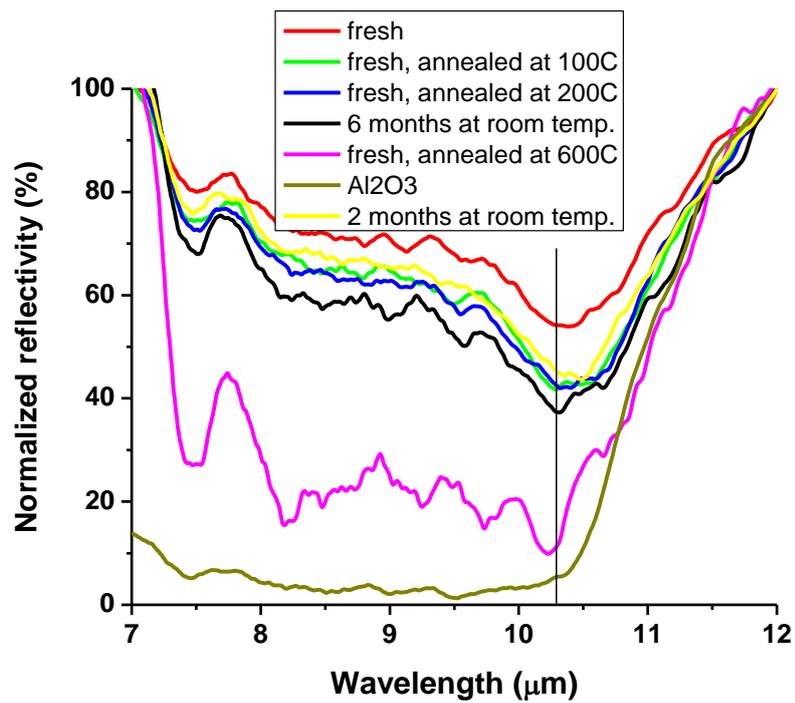

(b)

Fig. 2

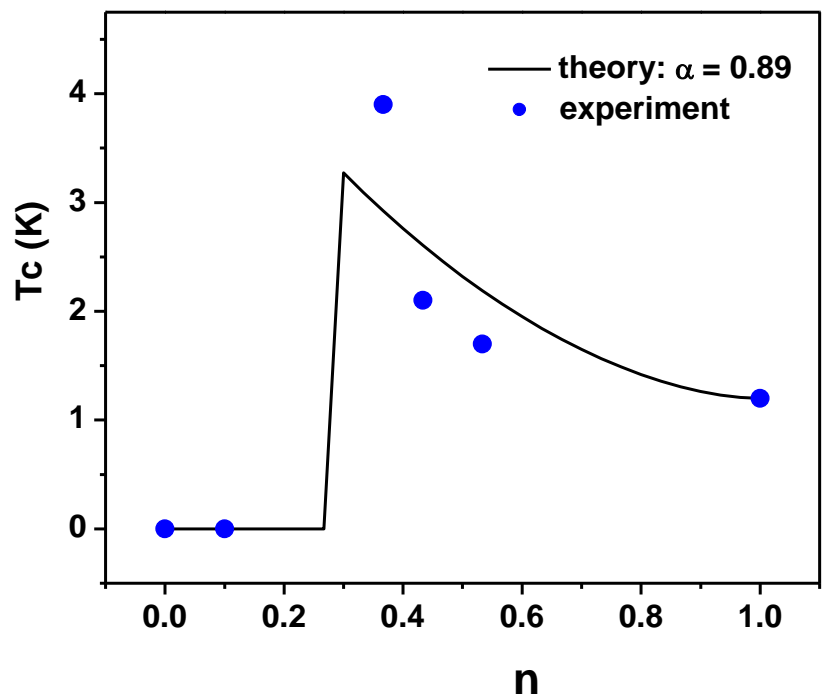

(a)

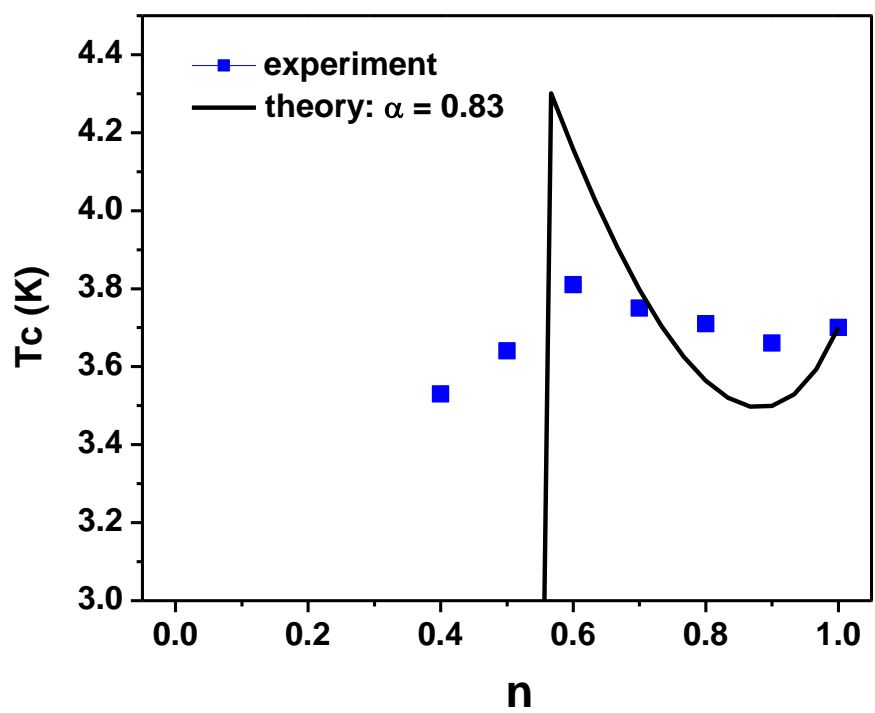

(b)

Fig. 3

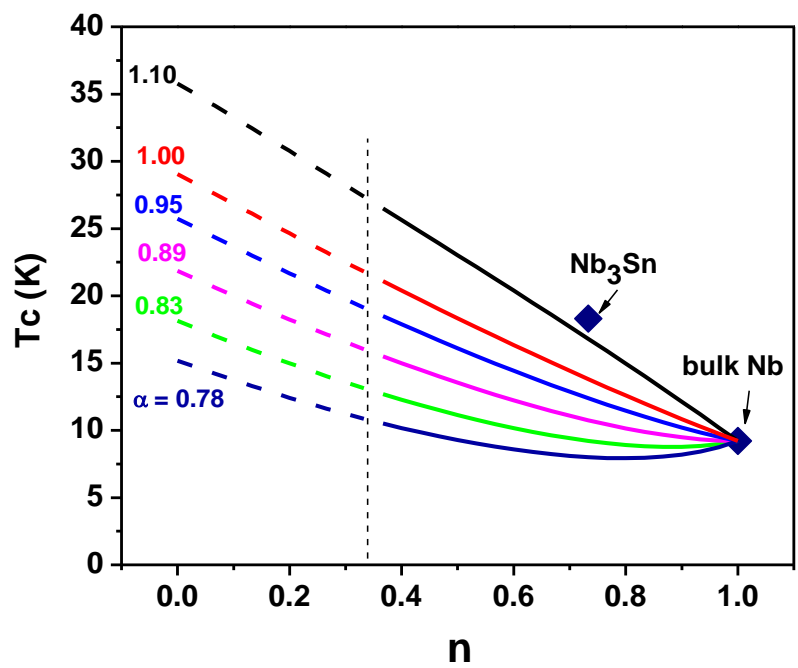

(a)

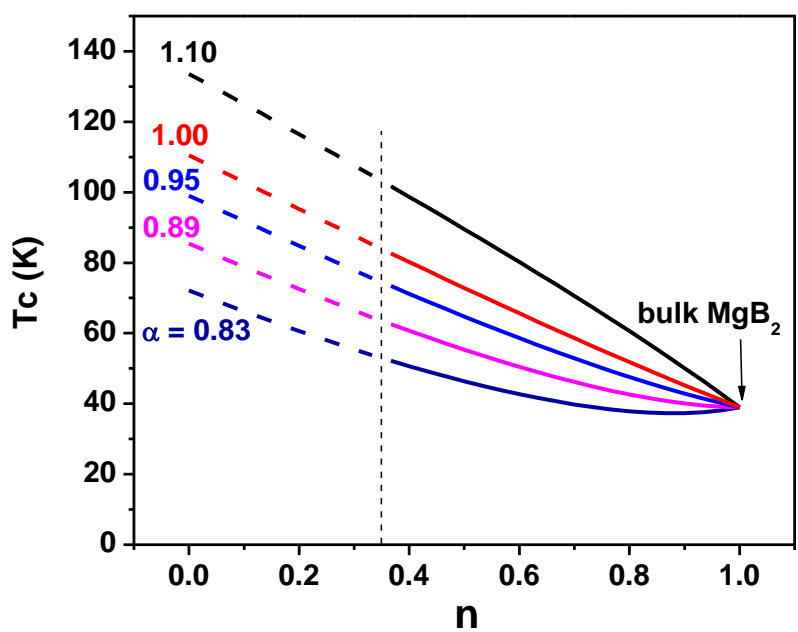

(b)

Fig. 4